\documentclass[twocolumn,pre]{revtex4}

\usepackage{graphicx}
\usepackage{dcolumn}
\usepackage{bm}

\usepackage{graphics}
\begin{document}

\title{Theory of DNA translocation through
narrow ion channels and nanopores with charged walls}

\author{Tao Hu}
\author{B. I. Shklovskii}
\affiliation{Theoretical Physics Institute, University of
Minnesota, Minneapolis, Minnesota 55455}

\date{\today}

\begin{abstract}

Translocation of a single stranded DNA through genetically
engineered $\alpha$-hemolysin channels with positively charged walls
is studied. It is predicted that transport properties of such
channels are dramatically different from neutral wild type
$\alpha$-hemolysin channel. We assume that the wall charges
compensate the fraction $x$ of the bare charge $q_{b}$ of the DNA
piece residing in the channel. Our prediction are as follows (i) At
small concentration of salt the blocked ion current decreases with
$x$. (ii) The effective charge $q$ of DNA piece, which is very small
at $x = 0$ (neutral channel) grows with $x$ and at $x=1$ reaches
$q_{b}$. (iii) The rate of DNA capture by the channel exponentially
grows with $x$. Our theory is also applicable to translocation of a
double stranded DNA in narrow solid state nanopores with positively
charged walls.

\end{abstract}

\maketitle

\section{Introduction}

A DNA molecule in a water solution carries negative charge. With the
help of applied voltage $V$, it can translocate through an ion
channel located in a lipid membrane or through a solid state
nanopore in a semiconductor film. In this paper we are interested in
the cases when DNA barely fits into a narrow pore leaving only a small
gap for water with the width $d < l_B$, where $l_B = e^{2}/\kappa k_B T$
is the Bjerrum length and $\kappa$ is the dielectric constant of water.
An intensively studied example is the translocation of a single
stranded DNA (ssDNA) molecule through an $\alpha$-hemolysin
($\alpha$-HL) channel~
\cite{Henrickson,Meller2001,Meller2002,Meller2003,Sauer,Mathe,Ambjörnsson,Nakane,Bonthuis}.
With the average internal diameter $\sim 1.7\,$nm the channel can accomodate
ssDNA molecule with $\sim 1\,$nm diameter. In this case, $d =0.35~$nm$ <l_B$.

Our theory also should be applicable to a double helix DNA (dsDNA)
with $\sim 2\,$nm diameter translocating through a narrow solid
state nanopore with $2a \sim 3\,$nm diameter~\cite{Li,Min}. On one
hand, no experimental data is available for that narrow nanopores.
On the other hand, there is impressive progress in making and
studying wider nanopores~\cite{Li,Min,Lemay,Aks,Aks1}.

\begin{figure}[htb]
\centering
\includegraphics[width=0.45 \textwidth]{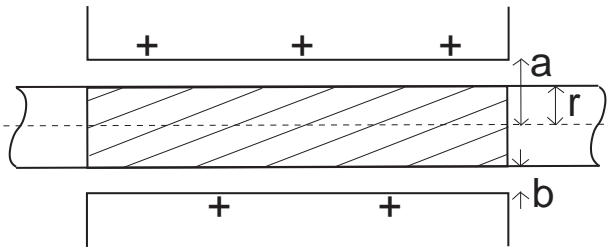}
\caption{The cross-section of the membrane and the channel with
radius $a$. Positive charges of the internal wall of the channel
are shown schematically. The captured DNA is shown as a cylinder
with radius $r$. A piece of DNA fitting into the channel is
shaded. DNA phosphates of this piece have the total charge $q_b =
-12e$. They are neutralized by the wall charges $x|q_b|$ and the
charge of mobile cations $(1-x)|q_b|$.} \label{figdna}
\end{figure}

The peculiarity of narrow channels is related to the fact that the
dielectric constants of the channel stem and lipids and the
dielectric constant of the body of DNA are much smaller than
dielectric constant of water. When the water filled gap between DNA
and the channel wall is narrow ($d < l_B$), the electric field of
small ions is squeezed in the gap.  Potential of interaction of
charges becomes logarithmic. This creates the electrostatic barrier
for the ion current similar to one which was intensively studied for
the ion transport through narrow DNA free
channels~\cite{Parsegian,Zhang,Kamenev}.

Previous discussion of the role of this barrier for ion transport
in the case of DNA translocation~\cite{Jing} was narrowly focused
on neutral channels, because the wild type $\alpha$-HL channel can
be considered practically neutral. A good measure of neutrality of
a channel is its cation/anion selectivity, measured by the ratio
of cation to anion currents. For the wild type $\alpha$-HL channel
this ratio is equal 1.1, while it is equal to unity for a exactly
neutral channel. It is known that at pH 7 the $\alpha$-HL channel
is neutral in the body of stem, but has the ring of 7 negative
charges near the bottom of the the narrow cylindrical part of the
channel (stem)~\cite{HLimage}. These charges are screened by the
salt in the water outside the stem and, therefore, do not
determine the transport through the channel which remains weakly
selective. Solid state nanopores can be neutral, too. For a
neutral narrow channel Ref.~\cite{Jing} addressed several
challenging problems posed by the experimental
data~\cite{Henrickson,Meller2001,Meller2002,Meller2003,Sauer,Mathe,Ambjörnsson,Nakane,Bonthuis}.

First, Ref.~\cite{Jing} explained how the electrostatic barrier
makes the blocked by DNA current $I_b$ at least 10 times smaller
than the open current $I_{0}$. Second, the effective stall charge
$q_s$ of the piece of DNA residing in the channel was calculated.
This charge determines the force $F_{s}= q_s V/L$ stalling DNA
against the voltage $V$ ($L$ is the length of the channel). It was
shown that for a neutral channel with small $I_b /I_{0}$ the
charge $q_s \simeq q_b I_{b}/I_{0}$, where $q_{b}$ is the bare
charge of the piece of DNA occupying the channel (for $\alpha$-HL
channel $q_{b} = -12e$). In agreement with experiments this
results in very small absolute value of the stall charge, namely
$q_s \sim -1e$. Third, the origin of an the exponentially small
and growing with the salt concentration DNA capture rate was
elucidated.

Recently, genetically modified $\alpha$-HL channels became
available~\cite{Aka,Mer}. In this paper we concentrate on those of
them, which have amino acids with positive residues on the internal
wall of the narrow cylindrical part of the channel (stem). Internal
walls of solid state nanopores may also be charged. The charge
density of these walls can be tuned by different chemical treatments
or just by a change of the solution pH. Thus, our theory for charged
$\alpha$-HL channels simultaneously addresses narrow charged
nanopores used for dsDNA translocation experiments~\cite{Min}.

We assume below that the fraction $ x $ of the bare charge $q_b$
of DNA piece fitting into the channel is compensated by positive
internal wall charges, which are roughly speaking randomly
distributed on internal wall of the channel (Fig.~\ref{figdna}).
We predict below that this simple assumption leads to a number of
dramatic changes of DNA translocation in comparison with a neutral
channel. Let us list these predictions:

(i) The blocked ion current $I_b$ becomes even smaller than in the
neutral channel particularly at small concentrations of salt.

(ii) The effective charge $q_s$ of the piece of DNA residing in the
channel grows with $x$ as $q_s \simeq xq_{b}$. At $x = 1$ the stall
and bare charges of DNA are almost equal. The large effective charge
will makes possible DNA manipulation with the help of small
voltages.

(iii) The barrier for the DNA capture decreases with $x$. As a
result the DNA capture rate exponentially grows with $x$ and the
number of translocation events observed in a given experiment
increases. This should lead to much more effective averaging of the
noise and may prove helpful in attempts of DNA sequencing. At some
$x = x_c$ the capture barrier vanishes. At $x_c < x \leq 1$ DNA is
attracted to the channel. The capture rate then is only diffusion
limited and independent on $x$. On the other hand, for a captured
DNA the probability to escape from the channel becomes activated.
The escape barrier grows with $x$ at $x > x_c$.

The structure of our paper is simple. It consists of three
sections leading to conclusions (i), (ii) and (iii) respectively.

\section{Release of counterions and blocked ion current}
\label{sec_Electrostatics}

In the case of $\alpha$-HL channel we assume the ssDNA molecule is a
rigid cylinder coaxial with the channel. The inner radius of the
$\alpha$-HL channel is $a\!\simeq\! 0.85\,$nm, and the radius of the
ssDNA molecule is $r\!\simeq\! 0.5\,$nm.
Salt ions are located in the water-filled gap between them, with
thickness $b\!\simeq\! 0.35\,$nm. The length of the channel is
$L\!\simeq\! 5\,$nm. This kind of model is even more appropriate for
double helix DNA in a wider (say $4\,$nm in diameter) cylindrical
solid state nanopore \cite{Li,Min}.

The dielectric constant of the channel or the ssDNA molecule
($\kappa'\!\sim\! 2$) is much smaller than that of water
($\kappa\!\simeq\! 80$). So if ssDNA is neutralized by cations and
there is an extra charge $e$ in the thin water-filled gap between
the channel internal wall and ssDNA, the electric field lines
starting from this charge are squeezed in the gap. This results in
a high self energy of the charge~\cite{Parsegian,Zhang,Jing}.
According to the estimate of Ref~\cite{Jing} for the case of ssDNA
in $\alpha$-HL channel, the self energy of the charge in the
middle of the channel is $\sim 5 k_{B}T$. Here and everywhere in
this paper $T$ is the room temperature.

Because of the large self energy of a charge in the narrow water
gap, the piece of ssDNA inside the wild type neutral $\alpha$-HL
channel is neutralized by counterions, say K$^{+}$ in KCl
solution~\cite{Jing}. ssDNA covered by cations presents a conducting
DNA backbone wire responsible for the blocked ion current $I_b$ at
small concentration of salt $c < 1$M. In this range of
concentrations $I_b$ is practically $c$ independent~\cite{Bonthuis}.
At larger concentration $c\geq 1$M additional pairs of anions and
cations in the channel provides a parallel to DNA backbone wire
mechanism of conductivity. (Recall that DNA backbone wire occupies
only a small fraction of the water filled gap.)  The linear growth
of $I_b$ with $c$ at $c\geq 1$M is an experimental evidence for the
second mechanism of conductivity~\cite{Bonthuis}

In a mutated positively charged channel situation is rather
different. Let us consider the channel with uniformly distributed 12
positive charges ($x = 1$). We argue that in this case both ssDNA
and internal wall charges release their counterions into the
surrounding salt solution. The net charge of the channel is still
zero and, thus, there is practically no price in the Coulomb energy.
On the other hand, counterion release leads to the large gain in
their entropy. As a result DNA backbone wire looses its carriers and
becomes an insulator. Therefore, $I_b$ is determined only by
contribution of additional pairs of salt ions. This should lead to
linear dependence of $I_b$ on $c$ in the whole range of salt
concentrations. In other words, $I_b$ becomes much smaller than in
the wild type channel at small $c < 1$M, but is not strongly changed
at larger concentration of salt.

So far we talked about wall charges totally compensating the bare
charge of DNA ($x = 1$ ). At $x < 1$ DNA counterions are only
partially released and conductance of DNA backbone wire is only
partially depleted. Although the number of counterions on the DNA
wire is proportional to $x$, their mobility may somewhat grow with
decreasing $x$ due to the increase of the number of empties. It is
possible, but seems unlikely that this growth leads to the
conductance maximum at $x \sim 0.5$.

\section{Effective charge of DNA}
\label{sec_Effectcharge}

As we mentioned above for the wild type channel the stall charge
of DNA $q_s$ is much smaller than the bare charge of DNA $q_b$.
Let us remind why this happens. Counterions neutralizing DNA in
the channel receive from electric field momentum with the same
absolute value as DNA, but in the opposite direction. Most of the
time counterions are bound to DNA charges and transfer all
received momentum to DNA. During this time, the net electric field
force acting on DNA vanishes. At rare moments when counterions get
free and move along the channel contributing to $I_b$, they
transfer half of their momentum to the internal channel wall. This
deficit of momentum transfer to DNA results in a small net average
force on DNA and its small effective charge $q_s$~\cite{Jing}.

In a channel where positively charged walls compensate the bare
charge of DNA, the balance of forces is completely different. When
countrerions of DNA and walls are released, electric field
provides opposite momentums to DNA and to the wall charges. The
latter are static and, therefore, transfer all their momentum to
the wall. Thus, DNA gets its momentum only directly from electric
field. This means that $q_s = q_b$.

So far we talked about the channel which totally compensates charges
of DNA ($x = 1$). Similar logic leads to the result $q_s = xq_b$ for
any $x < 1$.

\section{DNA Capture and Escape rates}
\label{sec_Capturerate}

Besides the blocked current and the stall charge one can measure the
average time between the two successive translocation events,
$\tau$, or the capture rate $R_c = 1/\tau$ of a DNA molecule into
the channel. It is natural to compare the observed value of $R_c$
with the diffusion limited rate $R_D$ of ssDNA capture. For the wild
type neutral $\alpha$-HL channel this comparison shows that $R_c \ll
R_D$. The capture rate at zero voltage $R_{c}(0)$ is so small that
all experiments are actually done with a large applied voltage $V =
50 - 200$ mV. Apparently there is a large barrier for the DNA
capture. A large part of this barrier is due to the loss of the
conformational entropy of ssDNA. The capture barrier, however,
depends on the salt concentration, what means that a part of it has
an electrostatic origin. The reason for such an electrostatic
barrier is as follows~\cite{Jing}. When a DNA molecule enters the
channel, the DNA counterions are squeezed in the narrow water-filled
space surrounding the DNA. Due to this compression the total free
energy of DNA and ions is higher for DNA in the channel than for DNA
in the bulk. In agreement with experiment this barrier decreases
with growing $c$ because the entropy of counterions in the bulk
solution decreases and, therefore, the price for compression is
smaller.

In the case of a channel with positively charged walls the ssDNA
does not need to bring all its counterions into the channel, because
there are already some positive charges. Thus, the charge $xq_b$ is
released by DNA to the bulk of solution making the electrostatic
barrier for DNA smaller. Additional, roughly speaking, equal gain is
provided by release of counterions of the wall charges, which screen
the walls in the absence of DNA. Thus, due to the counterion release
the electrostatic barrier becomes $1-2x$ times smaller. At $x =1/2$
the electrostatic barrier vanishes, but the conformation barrier
remains intact and the capture rate is still activated. At $x > 1/2$
the electrostatic contribution to the total barrier becomes negative
and at small enough concentration of salt, when $x=x_c<1$, it
eventually compensates the conformation barrier, so that $R_c =
R_D$. At $x>x_c$ the capture rate saturates at $R_D$, but the escape
rate has an activation energy. The linear dependence of the barrier
on $x$ can be measured.

To summarize, in this paper we studied DNA translocation through
narrow channels with positively charged walls. Our predictions for
the stall effective charge and capture rate are dramatically
different from the case of neutral channels. Interpretation of the
stall effective charge theory also becomes much simpler. This
paper extends theory of narrow channels started in
Ref.~\cite{Jing}. Meanwhile, the detail, theoretical description
of wider channels became available~\cite{Lemay}. Therefore, one
can ask when our theory crossovers to results~\cite{Lemay}. The
crossover happens when the width of water gap $d$ becomes larger
than $l_B$. For double helix DNA this happens for nanopores with
diameter $3.5~$nm or larger.

We are grateful to A. Aksimentiev, T. Butler, J. Gundlach,
Jingshan Zhang, J. J. Kasianovich, O. V. Krasilnikov and A. Meller
for useful discussions.


\end{document}